\documentclass[aps,prl,twocolumn,superscriptaddress,showpacs]{revtex4}
\usepackage{graphicx}

\usepackage[latin1]{inputenc}
\usepackage{bm}% bold math

\begin{document}

\renewcommand{\vec}[1]{{\mathbf #1}}

\title{Thermal signatures of Little-Parks effect in the heat capacity
of mesoscopic superconducting rings}
\author{Florian R. Ong}
\email[]{florian.ong@grenoble.cnrs.fr}
\affiliation{
Centre de Recherches sur les Tr\`{e}s Basses Temp\'{e}ratures, CNRS and
Universit\'{e} Joseph Fourier, 38042 Grenoble, France}
\author{Olivier Bourgeois}
\email[]{olivier.bourgeois@grenoble.cnrs.fr}
\affiliation{
Centre de Recherches sur les Tr\`{e}s Basses Temp\'{e}ratures, CNRS and
Universit\'{e} Joseph Fourier, 38042 Grenoble, France}
\author{Sergey E. Skipetrov}
\affiliation{Laboratoire de Physique et Mod\'elisation des Milieux Condens\'es,\\
Maison des Magist\`{e}res, CNRS and Universit\'{e} Joseph Fourier, 38042
Grenoble, France}
\author{Jacques Chaussy}
\affiliation{
Centre de Recherches sur les Tr\`{e}s Basses Temp\'{e}ratures, CNRS and
Universit\'{e} Joseph Fourier, 38042 Grenoble, France}

\date{\today}

\begin{abstract}

We present the first measurements of thermal signatures of the Little-Parks effect using a highly sensitive nanocalorimeter. Small variations of the heat capacity $C_p$ of 2.5 millions of non interacting micrometer-sized superconducting rings threaded by a magnetic flux $\Phi$ have been measured by attojoule calorimetry. This non-invasive method allows the measurement of thermodynamic properties --- and hence the probing of the energy levels --- of nanosystems without perturbing them electrically. It is observed that $C_p$ is strongly influenced by the fluxoid quantization (Little-Parks effect) near the critical temperature $T_c$. The jump of $C_p$ at the superconducting phase transition is an oscillating function of $\Phi$ with a period $\Phi_0=h/2e$, the magnetic flux quantum, which is in agreement with the Ginzburg-Landau theory of superconductivity.

\end{abstract}


\maketitle

% ***********************************************************

Little and Parks \cite{LPPRL9,PLPR133} were the first to demonstrate that the
superconducting transition temperature of a hollow cylinder oscillates with the magnetic flux $\Phi$
threading it, the period of the oscillations being the magnetic flux quantum $\Phi_0=h/2e$.
More recently, a similar phenomenon was observed in doubly connected thin-film
mesoscopic loops \cite{MoshNature373}.
The quantization of the fluxoid --- the sum of the magnetic flux through a given surface and the circulation of screening supercurrents along the curve bounding this surface --- was shown to be at the origin of the Little-Parks oscillations
\cite{MoshNature373,TinkhamBook,GroffPR10}.
To our knowledge, all available studies of Little-Parks oscillations involved resistance measurements and
required \textit{invasive} probing leads that fundamentally limited the implications of results, especially for
nanostructured systems \cite{MoshNature373} that can be easily perturbed by any external connections.
In particular, surprising behavior can result from the strong influence of the biasing current \cite{VloePRL69,StrunkPRB57}. One way to get around this problem has been the use of the so-called ballistic Hall magnetometry \cite{geim2000}, but up to now this technique has never been applied to study the Little-Parks effect.

Recent experiments have shown that calorimetric and, more generally, thermal probing of electronic nanosystems can be an interesting and entirely \textit{noninvasive} alternative to magnetic methods \cite{roukes,BourgPRL94,fon05,gia06}. This new approach has the advantage of probing thermodynamic properties of the system directly and has a potential of
providing an irreplaceable information about energy levels of nanosystems and the nature of phase transitions or heat exchanges (including the single-phonon regime \cite{roukes}) in nanometer-sized samples. In this Letter we use the calorimetric approach to perform the first \textit{contact-free} study of Little-Parks effect in mesoscopic
superconductors. We demonstrate that the heat capacity of a mesoscopic thin-film superconducting ring is strongly affected by the fluxoid quantization. In a ring the jump $\Delta C_p$ of heat capacity at the second-order superconducting-to-normal (SN) phase transition exhibits an oscillatory behavior with $\Phi$, similarly to what has been predicted theoretically for an infinitely long superconducting cylinder \cite{FinkPRB23}. The
concavity of the curve $\Delta C_p(\Phi)$ is opposite to that expected for a cylinder --- a difference that we are
able to explain using the Ginzburg-Landau (GL) theory of superconductivity.
Our results constitute the first direct experimental demonstration of modification of system behavior
at the second-order SN phase transition due to the reduced sample size. The magnitude of this modification can,
in addition, be modulated by changing an external parameter (magnetic field).
Such a behavior could be expected in many small system, as for example superconducting nano-grains, but has never been evidenced before.

\begin{figure}[h]
\includegraphics[width=7cm,angle=0]{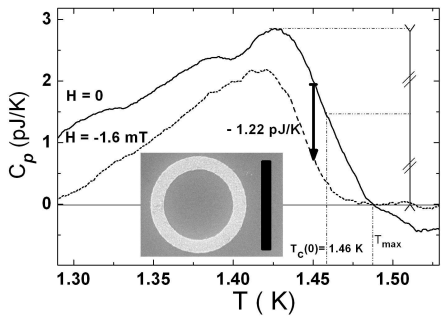}
\caption{\label{fig1}
Heat capacity versus temperature for magnetic fields $H = 0$ (solid line) and $H = -1.60$ mT (dashed line). The superconducting to normal phase transition, which is spread over 50 mK, starts when $C_p(T)$ starts to decrease and ends at the first kink of the drop ($T = T_{\rm max}$ for $H = 0$). The critical temperature $T_c(H)$ is defined as the temperature at a half-height of the transition. The inset shows an electron micrograph of a single loop: the scaling bar is 1 $\mu$m long.}
\end{figure}

Our calorimeter \cite{FominRSI68,BourgPRL94} is made of a thin ($\approx$ 5 $\mu$m) suspended silicon membrane (4 mm$\times$4 mm) integrating two thin film transducers: a copper heater and a NbN thermometer. The space between the
heater and the thermometer is used to replicate the nano-object (ring) to be studied as many times as possible, thus enabling the heat capacity signal of interest to rise out of noise. Our experimental method is based on scanning the heat capacity versus temperature or the applied magnetic field and measuring the heat capacity by ac-calorimetry \cite{BourgPRL94,SullivanPR173}: the temperature of the calorimeter membrane is forced to oscillate with an amplitude of $\approx$ 12 mK, allowing a sensitivity $\delta C_{\rm tot}/C_{\rm tot} \sim 5 \times 10^{-5}$ at 1 K, where $C_{\rm tot}$ is the heat capacity of the whole sample (nano-objects $+$ membrane).
In the present paper we study an array of $N = 2.47\times 10^6$ aluminum rings (see the inset of Fig.\ \ref{fig1}) with respectively outer and inner diameters $D = 1.10$ $\mu$m and $D_0 = 924$ nm,
arm width $w = 176$ nm, thickness $d =30$ nm, total mass of the array 97 ng. The rings are deposited on the calorimeter membrane by thermal evaporation after patterning a monolayer of PMMA resist by e-beam lithography. The separation of 1.5 $\mu$m between the centers of neighboring rings ensures that their mutual magnetic interaction is negligibly weak
\cite{BourgPRL94}, so that the heat capacity of the ensemble of rings is simply a sum of $N$ identical individual contributions. 
The calorimeter is cooled down in a $^3$He fridge ($T = 0.5$ K to 10 K) equipped with a superconducting coil supplying a magnetic field $\vec{H}$ normal to the plane of the rings.

Because the superconducting contribution to the heat capacity $C_p(T,H)$ is the only contribution to the total
measured signal $C_{\rm tot}(T,H)$ depending on the magnetic field,
we determine it as $C_{\rm tot}(T,H) - C_{\rm tot}(T,H^{\prime} > H_c(T))$ with $H_c(T)$ the critical
magnetic field at temperature $T$. Figure \ref{fig1} shows $C_p(T,H = 0)$ and $C_p(T,H = -1.60\rm~mT)$. 
The SN transition takes place at $T \approx 1.46$ K and has finite width $\sim 50$ mK. The latter can be due to different issues: to the reduced volume of a single ring \cite{MuhlPRB6,ZallyPRL27}, to a spread in the geometrical parameters of the rings that we have measured to be below 3\%, to the microscopic disorder, and also to the finite amplitude of temperature oscillations (12 mK) in our experiment, which is not negligible compared to the width of the transition. 
Figure \ref{fig1} explains how we measure the two important parameters of the SN transition in a constant magnetic field: the critical temperature $T_c(H)$ and the discontinuity $\Delta C_p(H)$ of the heat capacity.
The critical temperature at zero field
$T_c(H = 0) = 1.46$ K corresponds to a Pippard coherence length $\xi_0 = 1.9$ $\mu$m. The elastic
mean free path $l_e$ of the aluminum film was independently measured to be $l_e = 24$ nm. As $\xi_0 \gg l_e$, our
sample is in the dirty limit, and hence the zero temperature coherence length is given by $\xi(0)=0.85\sqrt{\xi_0 l_e} = 182$ nm.

\begin{figure}[ht]
\includegraphics[width=7cm,angle=0]{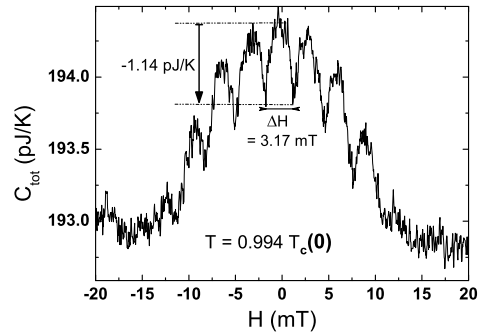}
\caption{\label{fig2}
Total heat capacity $C_{\rm tot}(H)$ of the sample (aluminum rings + addenda) at $T = 0.993 T_c(0) = 1.45$ K in decreasing magnetic field $H$ perpendicular to the plane of the rings. The variations of $C_{\rm tot}$ with $H$ are due only to the rings, since the response of the addenda is independent of $H$. The period $\Delta H = 3.17$ mT of oscillations of
$C_{\rm tot}(H)$ corresponds to a quantum of flux $\Phi_0$ through a circle of diameter $D_{\mathrm{eff}} = 910$ nm. The arrow shows the amplitude of $C_{\rm tot}$ variations
when sweeping the field from 0 to $H = -1.60$ mT, and has to be compared to the size of the arrow on Fig.\ \ref{fig1}.}
\end{figure}

We now perform $C_{\rm tot}(H)$ scans for many different but fixed temperatures $T$ near the critical region $T = (1.00 \pm 0.05)T_c(H=0)$. Since the heat capacity of the addenda is independent of $H$, only the aluminum rings contribute to the $H$-dependent part of $C_{\rm tot}(H)$. A typical scan is presented in Fig.\ \ref{fig2}. Starting at $H = 20$ mT, we slowly decrease the field by small steps and measure $C_{\rm tot}(H)$ at each $H$. Field cooled and zero field cooled scans show no difference. The large oscillations of $C_{\rm tot}(H)$ in the range $\left| H \right| < 15$ mT are
the major feature of Fig.\ \ref{fig2}. The period of these oscillations $\Delta H = (3.17 \pm 0.17)$ mT corresponds to one flux quantum $\Phi_0$ through a circle of diameter $D_{\rm eff} = (910 \pm 30)$ nm, which lies in between the inner and outer diameters of our rings.
The $\Phi_0$-periodic oscillations of the heat capacity of the array of superconducting rings is a direct signature of the Little-Parks effect. 

To highlight this aspect and to better expose the underlying physics, we will use the magnetic flux through a single loop $\Phi = H\pi D_{\rm eff}^2/4$ instead of $H$ to measure the strength of the applied magnetic field from here on. $C_p(\Phi)$ reaches local maxima at $\Phi = n \Phi_0$ ($n$ being an integer) and has sharp local minima at $n \Phi_0/2$. The amplitude of the oscillations of $C_p(\Phi)$ is about 1.14 pJ/K, which corresponds to 0.46 aJ/K $\approx 3\times 10^4 k_B$ per ring.
This value of amplitude is essentially built up by two distinct effects, as can be seen in Fig.\ \ref{fig1}, where the scans $C_p(T)$ are shown at two different values of $H$, corresponding to $\Phi = 0$ and $\Phi = -\Phi_0/2$. First,
we observe that $T_c(\Phi)$ decreases when $\left| \Phi \right|$ increases, thus yielding a smaller value of
$C_p$ at a given temperature. Second, the height $\Delta C_p(\Phi)$ of the heat capacity jump at $T_c(\Phi)$ is
smaller at $\Phi = -\Phi_0/2$ than at $\Phi = 0$. Both these features combine to give the measured value of the amplitude of heat capacity oscillations at a given temperature. 

\begin{figure}[h]
\includegraphics[width=7cm,angle=0]{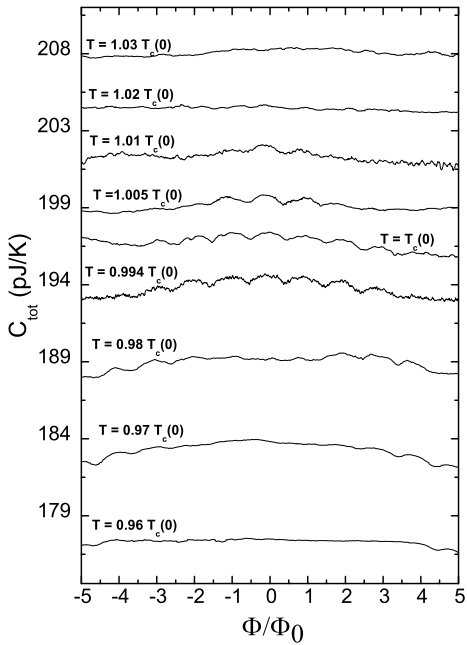}
\caption{\label{fig3}
Heat capacity $C_{\rm tot}$ of an array of superconducting rings as a function of magnetic flux $\Phi$ through a single ring for several fixed temperatures $T$, all close to $T_c(\Phi = 0) = 1.46$ K. For $T \leq 0.96 T_c(0)$ there are no oscillations: the superconductivity is monotonously suppressed as $\left| H \right|$ increases up to the critical field $H_c(T)$. For $0.97 T_c(0) \leq T \leq 0.98 T_c(0)$, oscillations appear at high fields, around the transition areas $\left| H \right| \approx H_c(T)$. For $0.99 T_c(0) \leq T < 1.01 T_c(0)$, the oscillations are spread over a large region around $\Phi = 0$, where 7 to 9 oscillations can occur. Above $T = 1.01 T_c(0)$, the oscillations get weaker and are present only in a close vicinity of $\Phi = 0$, until they completely disappear for $T > 1.02 T_c(0)$.}
\end{figure}

Figure \ref{fig2} can be understood in the same way as the resistance traces $R(H)$ obtained by the Little-Parks effect pioneers for superconducting cylinders \cite{LPPRL9,PLPR133,GroffPR10,MeyersPRB4}. The qualitative difference between $R(H)$ and $C_{\rm tot}(H)$ is that in the latter case the branches of parabolas that build up the curve are turned upside down.
It is important to emphasize that the observed oscillations of $C_{\rm tot}$ with $H$ for temperatures close to $T_c$ are not only due to the intrinsic dependence 
of the heat capacity of a superconducting ring on the applied magnetic field, but are also strongly affected
by the finite width of the SN transition. To illustrate this issue, we present $C_{\rm tot}(\Phi)$ for a set of different temperatures, all close to $T_c(\Phi = 0)$, in Fig.\ \ref{fig3}. 
Well below the critical temperature, i.e. for $T \leq 0.97 T_c(0)$, the oscillations of $C_{\rm tot}$ develop only at high magnetic fields approaching the critical field. As the temperature is increased, the oscillations become most pronounced around the zero magnetic flux point. If the temperature is raised above $T_c(0)$,
the interval of magnetic fields where the oscillations are present shrinks and finally vanishes for $T > 1.02 T_c(0)$.
The fact that the oscillations of $C_{\rm tot}$ with $\Phi$ are still present at $T$ slightly above $T_c(0)$ is due to the
finite width of the SN transition and to our way of defining $T_c$ (see Fig.\ \ref{fig1}).

\begin{figure}[h]
\includegraphics[width=7cm,angle=0]{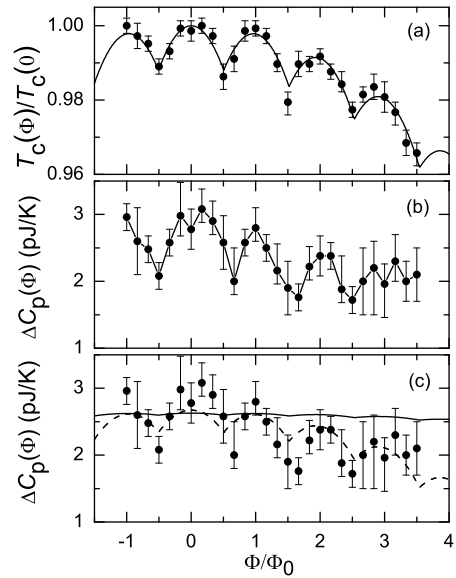}
\caption{\label{fig4}
(a). Experimental SN phase boundary of an aluminium ring (symbols). The error bars are due to the measurement noise and to the uncertainty when delimiting the transition region.  The solid line is a fit to the data with $\xi(0) = 98$ nm the only fit parameter.
(b). Heat capacity jump at the SN transition $\Delta C_p(\Phi)$ versus the magnetic flux $\Phi$ through a single ring. (c). The best fit to the data of panel (b) (dashed line) is obtained with $\xi(0) = 310$ nm; the solid line is for $\xi(0) = 98$ nm following from the fit in panel (a).}
\end{figure}

The ultimate manifestation of the Little-Parks effect in our thin-film superconducting rings can be obtained by measuring the dependence of the critical temperature $T_c$ of the SN transition and of the jump $\Delta C_p(\Phi)$ of heat capacity at the transition on the magnetic flux $\Phi$ through a single ring. Figure \ref{fig4}(a) shows the phase boundary diagram $T_c(\Phi)$ that we obtained by performing scans similar to those shown in Fig.\ \ref{fig1} at 28 different values
of $\Phi$. $T_c(\Phi)$ has been extracted from the position of the jump of heat capacity as illustrated in Fig.\ \ref{fig1}. The oscillations of $T_c$ with $\Phi$ seen in the figure are the manifestation of the Little-Parks effect
and have been previously observed by resistance measurements on superconducting cylinders \cite{GroffPR10,MeyersPRB4} and rings \cite{MoshNature373}.
Figure \ref{fig4}(a) is the first \textit{contact-free} measurement of the phase boundary $T_c(\Phi)$ because,
in contrast to previous studies, we do not attach leads to our sample.
The data of Fig.\ \ref{fig4}(a) can be compared to the phase boundary following from the Ginzburg-Landau theory of superconductivity \cite{TinkhamBook,TinkhamPR129}:
\begin{eqnarray}
T_c(\Phi) = T_c(0) \sqrt{\frac{1-\gamma}{1+\gamma}},
\label{tc}
\end{eqnarray}
where $\gamma = [2 \xi(0)/{\bar D}]^2
\{(n-\Phi/\Phi_0)^2 + [n^2/3 + (\Phi/\Phi_0)^2]$ $(w/{\bar D})^2\}$,
${\bar D} = (D + D_0)/2$ is the average ring diameter, $w \ll {\bar D}$ is assumed, and 
$n$ is an integer that maximizes $T_c$ at a given $\Phi$.
To arrive at Eq.\ (\ref{tc}), we find the difference $\Delta F$ between the free energies of the ring in the superconducting and normal states by solving the GL differential equation for the complex order parameter $\psi(\vec{r})$, neglecting variations of the latter in the radial and axial directions and assuming that the magnetic field inside the sample is equal to the applied field \cite{prepar}. These approximations are justified when the width $w$ of the ring is
smaller than the GL coherence length $\xi(T)$ and when the thickness $d$ of the ring is smaller than
both $\xi(T)$ and the magnetic field penetration depth $\lambda(T)$ \cite{bezryadin95,zhang97}, which
is the case in our experiments. $T_c(\Phi)$ is found from the requirement $\Delta F = 0$.
The best fit to our data with Eq.\ (\ref{tc}) is obtained with $\xi(0) = 98$ nm [the solid line in Fig.\ \ref{fig4}(a)] which is smaller than the value of $\xi(0)$ estimated from the measurements of the elastic mean free path.

The discontinuity $\Delta C_p(\Phi)$ of heat capacity at $T_c(\Phi)$ is presented on Fig.\ \ref{fig4}(b). Despite the large error bars taking into account the noise and the uncertainty when delimiting the SN transition region, $\Phi_0$-periodic oscillations of $\Delta C_p(\Phi)$ are clearly visible. $\Delta C_p(\Phi)$ exhibits sharp minima at $\Phi = n \Phi_0/2$, similarly to $T_c(\Phi)$.
The oscillatory behavior of $\Delta C_p(\Phi)$ also follows from the GL theory:
\begin{eqnarray}
\Delta C_p(\Phi) = -T \frac{\partial^2 \Delta F}{\partial T^2} = \Delta C_p(0) (1-\gamma)^{3/2} \sqrt{1+\gamma}.
\label{dc}
\end{eqnarray}
This equation provides a reasonable fit to the data with $\xi(0) = 310$ nm [see the dashed line in
Fig.\ \ref{fig4}(c)] which is 3 times larger than the value used in Fig.\ \ref{fig4}(a). Using $\xi(0) = 98$ nm
as in Fig.\ \ref{fig4}(a) yields a curve with the same periodicity, but with much less important amplitude
of oscillations [solid line in Fig.\ \ref{fig4}(c)].
Hence, the measured oscillations of $\Delta C_p$ with $\Phi$ appear more pronounced
than could be expected theoretically.
The reason for this discrepancy is not yet understood.

The oscillations of the discontinuity of heat capacity at the SN transition were studied theoretically for hollow,
infinitely long \textit{cylinders} \cite{FinkPRB23}. Although $T_c(\Phi)$ is given by the same Eq.\ (\ref{tc}) for both infinitely long cylinders and thin-film rings,
$\Delta C_p(\Phi)$ exhibits very different behavior in the two cases. Namely, for a cylinder the
parabolas constituting the $\Delta C_p(\Phi)$ curve in Fig.\ \ref{fig4}(c) would be turned upside-down \cite{FinkPRB23}.
This interesting difference between the ring and cylinder geometries comes from their very different flux
entry/expulsion properties and appear to be correctly captured by the GL theory.
This difference cannot be evidenced by simply measuring the SN phase boundary [Fig.\ \ref{fig4}(a)] because $T_c(\Phi)$
is expected to be the same for rings and cylinders.

In conclusion, we have applied a highly-sensitive and contact-free calorimetric method to study the Little-Parks effect in mesoscopic superconducting rings. We have shown that the heat capacity of the rings, the critical temperature of the superconducting (SN) transition, and the jump of the heat capacity at the transition are strongly modulated by the external magnetic field and exhibit clear signatures of fluxoid quantization. The amplitude of the heat capacity jump
$\Delta C_p$ at the SN transition oscillates with the magnetic flux $\Phi$ through a single ring with a period of the magnetic flux quantum $\Phi_0$. The observed concavity of the dependence of $\Delta C_p$ on $\Phi$
is dictated by the small (30 nm) thickness of the rings and is opposite to that expected for infinitely long
cylinders \cite{FinkPRB23}. 
Repeating the measurements reported above for smaller rings [ring diameter down to or smaller than $\xi(0)$] could
help to understand the peculiarities of Little-Parks in the so-called destructive regime \cite{LiuScience294,WangPRL95}.
More generally, our results suggest that modern nanocalorimetry has unprecedented and largely unexplored potentials
for the study of phase transitions in nanostructured systems.

We would like to thank E. Andr\'e, P. Lachkar, J-L. Garden, C. Lemonias, B. Fernandez, T. Crozes for technical support, P. Brosse-Maron, T. Fournier, Ph. Gandit and J. Richard for fruitfull discussions and help and the Institut de Physique de la Mati\`ere Condens\'ee (Grenoble) for financial support.

% **********************************************************

\end{document}